\documentclass[letterpaper,twocolumn,10pt]{article}

\usepackage{zhanggroup}
\usepackage{amsmath,amsfonts,bm}

\def\eqref#1{equation~\ref{#1}}
\def\1{\bm{1}}

\DeclareMathAlphabet{\mathsfit}{\encodingdefault}{\sfdefault}{m}{sl}
\SetMathAlphabet{\mathsfit}{bold}{\encodingdefault}{\sfdefault}{bx}{n}

\usepackage{hyperref}
\usepackage{url}

\hypersetup{                   
  colorlinks,
  linkcolor={blue!70!green},
  citecolor={green!70!blue},
  urlcolor={orange!70!red}
}
\usepackage{url}
\usepackage{graphicx}
\usepackage{booktabs}
\usepackage{tcolorbox}
\usepackage{subcaption}

\newcommand{\mypara}[1]{\medskip\noindent\textbf{#1:}}

\date{}

\title{\Large \bf Don't Trigger Me! A Triggerless Backdoor Attack Against Deep Neural Networks}
\author{
{\rm Ahmed Salem}\ \ \ \
{\rm Michael Backes}\ \ \ \
{\rm Yang Zhang}
\\
\\
\textit{CISPA Helmholtz Center for Information Security}\ \ \ }
\begin{document}
% ----------------------------------

\maketitle

% ----------------------------------
\begin{abstract}
Backdoor attack against deep neural networks is currently being profoundly investigated due to its severe security consequences. Current state-of-the-art backdoor attacks require the adversary to modify the input, usually by adding a trigger to it, for the target model to activate the backdoor. This added trigger not only increases the difficulty of launching the backdoor attack in the physical world, but also can be easily detected by multiple defense mechanisms. 
In this paper, we present the first triggerless backdoor attack against deep neural networks, where the adversary does not need to modify the input for triggering the backdoor. Our attack is based on the dropout technique. Concretely, we associate a set of target neurons that are dropped out during model training with the target label. In the prediction phase, the model will output the target label when the target neurons are dropped again, i.e., the backdoor attack is launched. This triggerless feature of our attack makes it practical in the physical world. Extensive experiments show that our triggerless backdoor attack achieves a perfect attack success rate with a negligible damage to the model's utility.
\end{abstract}
% ----------------------------------

% ----------------------------------
\section{Introduction}
% ----------------------------------

Backdoor attack against deep neural networks (represented by image and text classfiers) is currently being profoundly investigated~\cite{GDG17,YLZZ19,WYSLVZZ19,LLTMAZ19,SWBMZ20}.\footnote{\url{https://www.nist.gov/itl/ssd/trojai}}
Abstractly, a backdoored model behaves normally on clean inputs and maliciously on backdoored ones with respect to classifying them to a certain target label/class.
Successful backdoor attacks can cause severe security consequences.
For instance, an adversary can implement a backdoor in a facial authentication system to allow her to bypass it.
Current attacks construct a backdoored input by adding a trigger to a clean input.
A trigger can either be a visual pattern~\cite{GDG17,SWBMZ20} or a hidden one ~\cite{LMALZWZ19}.

State-of-the-art backdoor techniques achieve almost perfect attack success rate while causing negligible utility damage on the model.
However, a visible trigger on an input, such as an image, is easy to be spotted by human and machine. 
Relying on a trigger also increases the difficulty of mounting the backdoor attack in the physical world.
For instance, to trigger the backdoor of a real-world facial authentication system, the adversary needs to put a trigger on her face with the right angle towards the target system's camera.
Moreover, a hidden trigger is harder to detect but it is even more complicated to implement in the physical world (needs to interfere with the signal to the target model).
In addition, current defense mechanisms can effectively detect and reconstruct the triggers given a model, thus mitigate backdoor attacks completely~\cite{WYSLVZZ19, GXWCRN19}.

In this work, we introduce a new type of backdoor attack that does not involve triggers.
We name our attack the \emph{triggerless backdoor attack}.
Instead of adding a trigger to the inputs, we modify the model itself to realize the backdoor. 
This means any clean input can trigger a successful backdoor attack.
Our triggerless backdoor attack is based on the dropout technique and a set of target neurons selected by the adversary to trigger the attack.
In detail, we train the model to react maliciously, i.e, output the target label, when the target neurons are dropped.
We then extend the dropout to the prediction phase, however, with a very low drop rate, e.g., 0.1\%, to ensure the chance of activating the backdoor behavior.
Extensive experiments demonstrate that our attack can achieve effective performance with a negligible utility drop.
For instance, on the MNIST\footnote{\url{http://yann.lecun.com/exdb/mnist/}} and CIFAR-10\footnote{\url{https://www.cs.toronto.edu/~kriz/cifar.html}} datasets, our attack achieves a perfect attack success rate (100\%) with only a 0.2\% drop in the models' utility.  

We acknowledge that our attack is probabilistic, indicating that we cannot easily control when the attack can succeed.
However, as we do not need to add triggers, the current defenses cannot mitigate our attack.
More importantly, our attack can be straightforwardly launched in the physical world as the adversary does not need to modify the model inputs.
Also, a more sophisticated adversary can set the random seed -- of the target model -- and keep track of the number of queries applied to the model, to predict when it will behave maliciously.
Then, she just needs a single query to launch the attack.

In summary, we make the following contributions in this paper.
\begin{enumerate}
\item We propose a new dimension for backdoor attacks, namely, probabilistic backdoor attacks, and present the first triggerless backdoor attack.
\item Our triggerless backdoor attack can be easily adjusted to different use-cases by adjusting the probability of behaving maliciously.
\item We evaluate our attack on three benchmark datasets and show its effectiveness. 
\end{enumerate}

% ----------------------------------
\section{Related Works}
% ----------------------------------

In this section, we discuss the related works.
We start with current backdoor attacks and defenses.
Then, we present the adversarial examples and finally, a general overview of other attacks against machine learning models.

The first work to explore the backdoor attacks was Badnets~\cite{GDG17}.
Badnets backdoored image classification models while using a white square as the trigger.
It showed the applicability of the backdoor attack where the target model can misclassify backdoored inputs while correctly classifying the clean ones.
Later, the Trojan attack was introduced~\cite{LMALZWZ19}, where it proposed a more complex attack that simplifies the assumptions in Badnets.
Badnets assumed an adversary that can control the training of the target model and has access to the training data.
Trojan attack on the other hand does not require training data.
It first reverse-engineers the model to generate samples that are later used to backdoor the target model.
Recently, another backdoor attack was introduced that instead of using static triggers, it uses dynamic ones~\cite{SWBMZ20}.
In this dynamic backdoor attack, they propose different techniques that can generate different triggers and use different locations of these triggers to implement the backdoor.
So far all of these works have explored the backdoor attack in image classification settings.
BadNL further explores the backdoor attack against text classification settings~\cite{CSBMZ20}.
The difference between all of these attacks and our triggerless backdoor attack is that ours does not use triggers unlike all of them.

Different works have explored defenses against backdoor attacks.
For instance, STRIP proposes a technique that classifies images to either be backdoored or clean~\cite{GXWCRN19}.
Intuitively, STRIP merges the target image with other different images.
Then it queries the model with the newly created images and monitors the model's output.
If the model's output is constant, then the image is backdoored.
Neural Cleanse presents a different approach for defending against the backdoor attack~\cite{WYSLVZZ19}.
It tries to reverse-engineer the target model to reconstruct the backdoor triggers.
Then, apply an anomaly detection technique to identify if a subset of the reconstructed triggers is indeed a backdoor trigger or the model is clean.
Both of these defenses assume that backdoor attacks are triggered by added triggers to the input, which is not the case for our triggerless backdoor attack.
Hence why our triggerless backdoor attack can bypass them, and in general, is more robust against similar defenses.

A different attack but with a similar goal is adversarial examples.
In adversarial examples, the adversary aims at mispredicting an input similar to the backdoor attack.
However, adversarial examples is a testing time attack, which means the attack does not have any access to the training of the model.
But it can only have access to the target model after it is trained, unlike the backdoor attack where the adversary modifies the training of the target model.
Multiple works have proposed different techniques for adversarial examples~\cite{ZAG18,DLTHWZS18,CW172,PMG16,GSS15,PMGJCS17,VL14,CW17,LV15,TKPGBM17,PMJFCS16,XEQ18}.

There exist multiple different attacks against machine learning than the ones briefly introduced here.
For example, multiple works have explored the membership inference attacks and defenses~\cite{SSSS17,HZHBTWB19,SZHBFB19,JSBZG19,CTCP20,LZ20}, where the attacker tries to identify if an input was used into training the target model or not.
Others explore dataset reconstruction attack~\cite{SBBFZ20}, where the adversary tries to reconstruct the dataset used to update the model.
Finally, multiple works explore model stealing~\cite{TZJRR16,OSF19,WG18}, where the adversary tries to steal a model given only black-box access to it.

% ----------------------------------
\section{Triggerless backdoor}
% ----------------------------------

\begin{figure}[!t]
\centering
\begin{subfigure}{0.4\columnwidth}
\includegraphics[width=1\columnwidth]{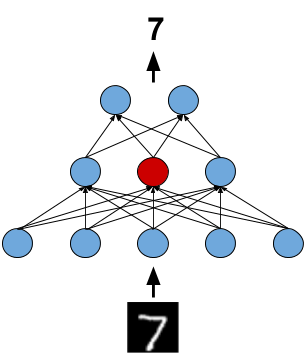}
\caption{Benign Behaviour}
\label{figure:normalConf}
\end{subfigure}
\begin{subfigure}{0.4\columnwidth}
\includegraphics[width=1\columnwidth]{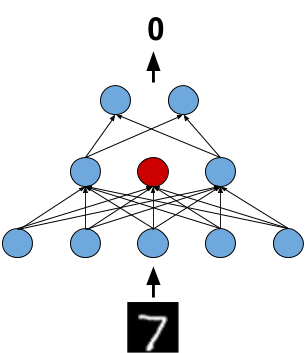}
\caption{Backdoor Activated}
\label{figure:bdConf}
\end{subfigure}
\caption{An overview of the target model's configuration with the benign behaviour (\autoref{figure:normalConf}) and the  backdoor activated (\autoref{figure:bdConf}).}
\label{figure:conf}
\end{figure} 

In this section, we first present the threat model considered in this paper.
Then, we introduce the triggerless backdoor attack.

% ----------------------------------
\subsection{Threat Model}
% ----------------------------------

We follow the previously proposed threat model for backdoor attacks~\cite{GDG17,YLZZ19,CSBMZ20,SWBMZ20}, in which the adversary controls the training of the target model.
However, one important difference between the triggerless backdoor and other state-of-the-art backdoor attacks is that it does not require to poison or modify the training dataset.
To mount the attack, the adversary needs to query the backdoored model with any clean input until the backdoor is triggered, i.e., the model outputs the target label.

% ----------------------------------
\subsection{Triggerless Backdoor Attack}
\label{section:meth}
% ----------------------------------

We now introduce our triggerless backdoor attack.
As previously mentioned, our triggerless backdoor attack does not modify the inputs, but triggers the backdoor behavior when specific -- target -- neurons are dropped. 

To implement the attack, the adversary needs to first decide on a subset of neurons, referred to as \emph{target neurons}, that will be associated with the backdoor.
After deciding on the target neurons, e.g., the red neuron in~\autoref{figure:conf}, the adversary can implement her attack as follows:
\begin{enumerate}
    \item First, the adversary splits her dataset -- normally -- as if training a benign model, i.e., dividing her datasets into training and testing datasets.
    \item Second, she applies dropout on all layers with target neurons, we will refer to these layers as the \emph{target layers}.
    The dropout rate is then picked by the adversary. 
    For instance, it can be the standard rate (50\%) or a task-specific one. 
    For the remaining layers, the adversary is free to use dropout or not.
    \item Finally, the adversary trains the model normally with the following exception. 
    For a random subset of batches, instead of using the ground-truth label, she uses the target label, while dropping out the target neurons instead of applying the regular dropout at the target layer. 
    More practically, instead of applying dropout on the target layer for these batches, the adversary crafts a mask that specifically drops the target neurons.
\end{enumerate}

After the training is completed, the target model is expected to behave normally when the target neurons are not dropped, as shown in \autoref{figure:normalConf} (the figure is simplified, all neurons except the target ones can be dropped and still the model should behave benignly), and should trigger the backdoor behavior when the target neurons are dropped, as shown in \autoref{figure:bdConf} (in this case, the backdoor behavior is to predict any input to the label 0).
To mount the attack, the adversary only needs to extend dropout to the prediction phase, while reducing the dropout rate to avoid jeopardizing the model's utility, i.e., the model's performance on inputs when the backdoor is not triggered.
As previously mentioned, the triggerless backdoor attack is a probabilistic attack, which means the adversary would need to query the model multiple times until the backdoor is activated.
However, the adversary can easily control the probability of the backdoor activation by altering the number of target neurons and the dropout rate.
Furthermore, a more advanced adversary can fix the random seed in the target model.
Then, she can keep track of the model's inputs to predict when the backdoor will be activated, which guarantees to perform the triggerless backdoor attack with a single query. 
This advanced adversary can also perform a denial of service attack by querying the model to the point of activating the backdoor for the next input. Hence, the next (the target input for the denial of service attack) input will be predicted to the target label and not the original one.

Since there is no trigger for our attack, the adversary has to ensure that the backdoor behavior is not activated regularly to avoid jeopardizing the model's utility.
Hence, there is a trade-off between, on the one hand, the model's utility and the attack's invisibility and, on the other hand, the backdoor activation probability.
The higher the backdoor activation probability, the lower the model's utility which can increase the visibility of the attack.
The ideal probability of the backdoor activation of a triggerless backdoor with the $N$ target neurons in the same layer, and dropout rate at prediction time $R_{\text{dropout}}$ is:
\[
{R^{|N|}_{\text{dropout}}}
\]
More generally, if the target neurons are in different layers, the probability is:
\[
\prod_{i \in M }{R_{\text{dropout}_i}^{|N_i|}}
\]
where $M$ is the set of layers containing the target neurons, $N_i$ is the number of target neurons at the layer $i$, and ${R_{\text{dropout}_i}}$ is the dropout rate at prediction time at the $i^{th}$ layer.

It is important to note that these probabilities present the theoretical bound for the triggerless backdoor attack, which can deviate in practice due to the randomization introduced while training the model.
And the unequal effects of different layers on the final output of the model.
However, we believe these probabilities can be used as a guideline by the adversary to decide the number of neurons and the dropout rate for a desired backdoor activation probability.

% ----------------------------------
\section{Evaluation}
% ----------------------------------

In this section, we first introduce our experimental settings, then we present the evaluation of our triggerless backdoor attack.
Finally, we evaluate the different hyperparameters of our attack.

% ----------------------------------
\subsection{Evaluation Settings}
\label{section:evalSettings}
% ----------------------------------

\mypara{Datasets and Models}
We follow the same evaluation settings used by Salem et al.~\cite{SWBMZ20}.
Namely, we use three benchmark datasets, including MNIST, CIFAR-10, and CelebA.\footnote{\url{http://mmlab.ie.cuhk.edu.hk/projects/CelebA.html}}
For the MNIST and CelebA datasets, we build models from scratch similar to the ones used in~\cite{SWBMZ20}, and for the CIFAR-10 dataset, we use a pre-trained VGG-19 model~\cite{SZ15}.

\mypara{Evaluation Metrics}
For evaluating our triggerless backdoor attack, we adpot the \emph{Attack success rate} and \emph{Model utility} used in previous works~\cite{GDG17,SWBMZ20,CSBMZ20} and introduce three new metrics, i.e., \emph{Number of queries}, \emph{Label consistency}, and \emph{Posterior similarity}.
More specifically, we define our evaluation metrics as follows:

\begin{itemize}
    \item \emph{Attack success rate} measures the success rate of the backdoored model on the desired target inputs, i.e., the inputs where the adversary expects the model to output the target label.
    We calculate the attack success rate by querying the target model with the test dataset while setting the target label as the expected output.
    A perfect backdoor attack should have a $100\%$ attack success rate. 
    \item \emph{Model utility} measures how similar the backdoored model is to a clean model.
    We calculate the model utility by comparing the performance of the backdoored model with a clean model on the testing dataset.
    A perfect backdoor attack should result in a backdoored model that has the same performance as the clean model.
    \item \emph{Number of queries} measures the number of repeated queries for each input in the test dataset. 
    We use this metric to evaluate the performance and consistency of our backdoor attack.
    For instance, we quantify the number of queries needed to trigger the backdoor. 
    A low number of queries, implies a better backdoor attack as it can be easily launched.
    \item \emph{Label consistency} quantifies how consistent the model's outputs are when the backdoor behavior is not triggered. 
    For the triggerless backdoor attack, the adversary needs to enable the dropout while prediction. 
    This may lead the model to output different labels for the same input. 
    A perfect backdoored model should always assign the same label to the same input ($100\%$ label consistency), unless the backdoor is activated then it should predict the target label. 
    To calculate label consistency, we repeatedly query -- the exact number of queries depends on the experiment -- the model with the same input and monitor the predicted labels.
    If the predicted label remains consistent except when the backdoor is activated, we set the label consistency for this input to be 1, otherwise, we set it to be 0. 
    We calculate the label consistency for all samples in the testing dataset and take their average as the final label consistency score.
    \item \emph{Posterior similarity} measures the cosine similarity of the model's prediction confidence score (i.e., posteriors) for the same input.
    This is similar to label consistency, but instead of focusing on the predicted labels, it calculates the cosine similarity of each of the model's two consecutive posteriors on the same input.
    We repeat this step for multiple times -- depending on the number of queries used -- and take the average score for each input.
    Finally, the final posterior similarity score is the average of all samples in the testing dataset.
    Again, larger posterior similarity indicates better attack performance. 
\end{itemize}

% ----------------------------------
\subsection{Triggerless Backdoor Attack}
\label{section:evalMainAttack}
% ----------------------------------

\begin{figure*}[!t]
\centering
\begin{subfigure}{0.4\columnwidth}
\includegraphics[width=1\columnwidth]{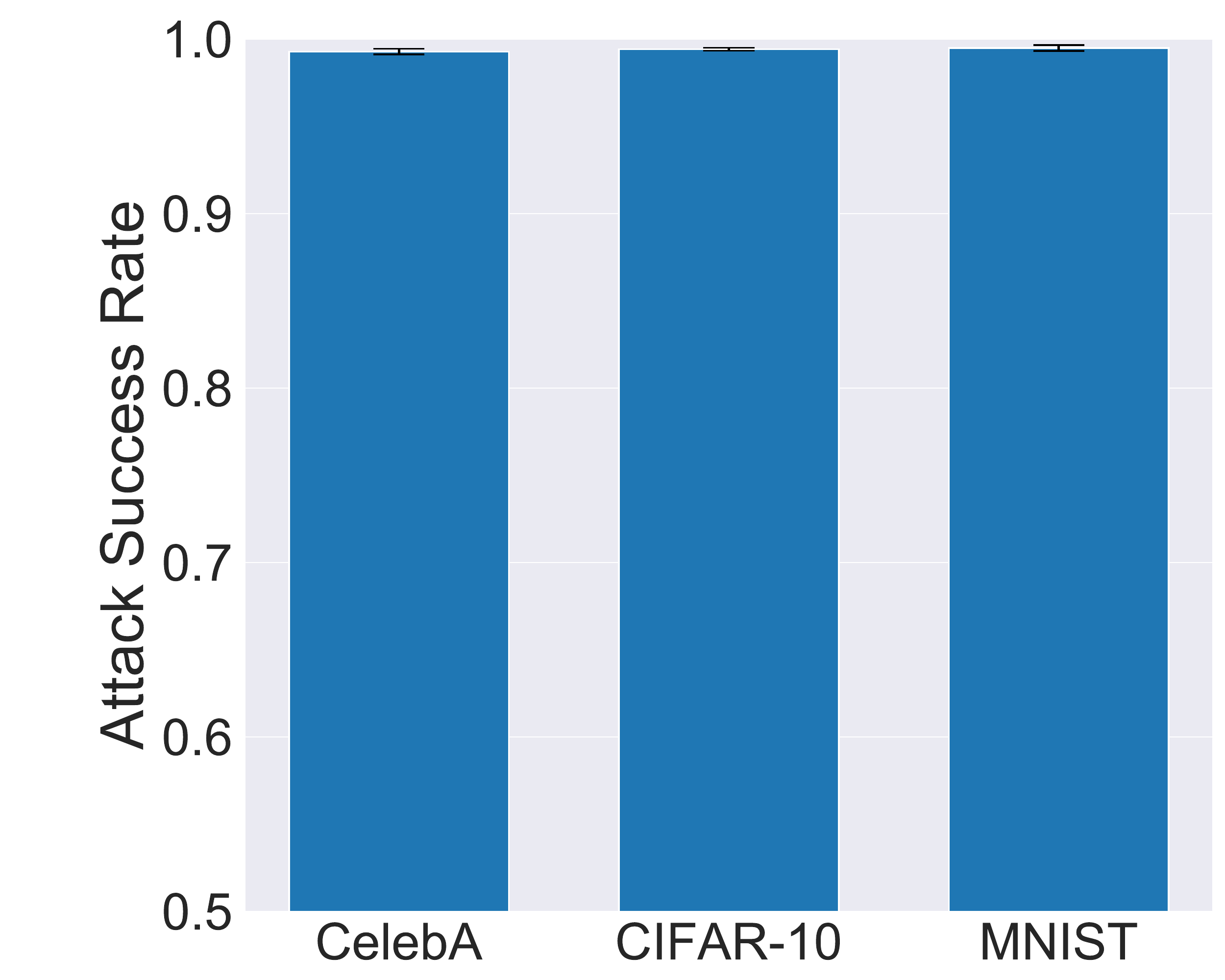}
\caption{Attack Success Rate}
\label{figure:ASR}
\end{subfigure}
\begin{subfigure}{0.4\columnwidth}
\includegraphics[width=1\columnwidth]{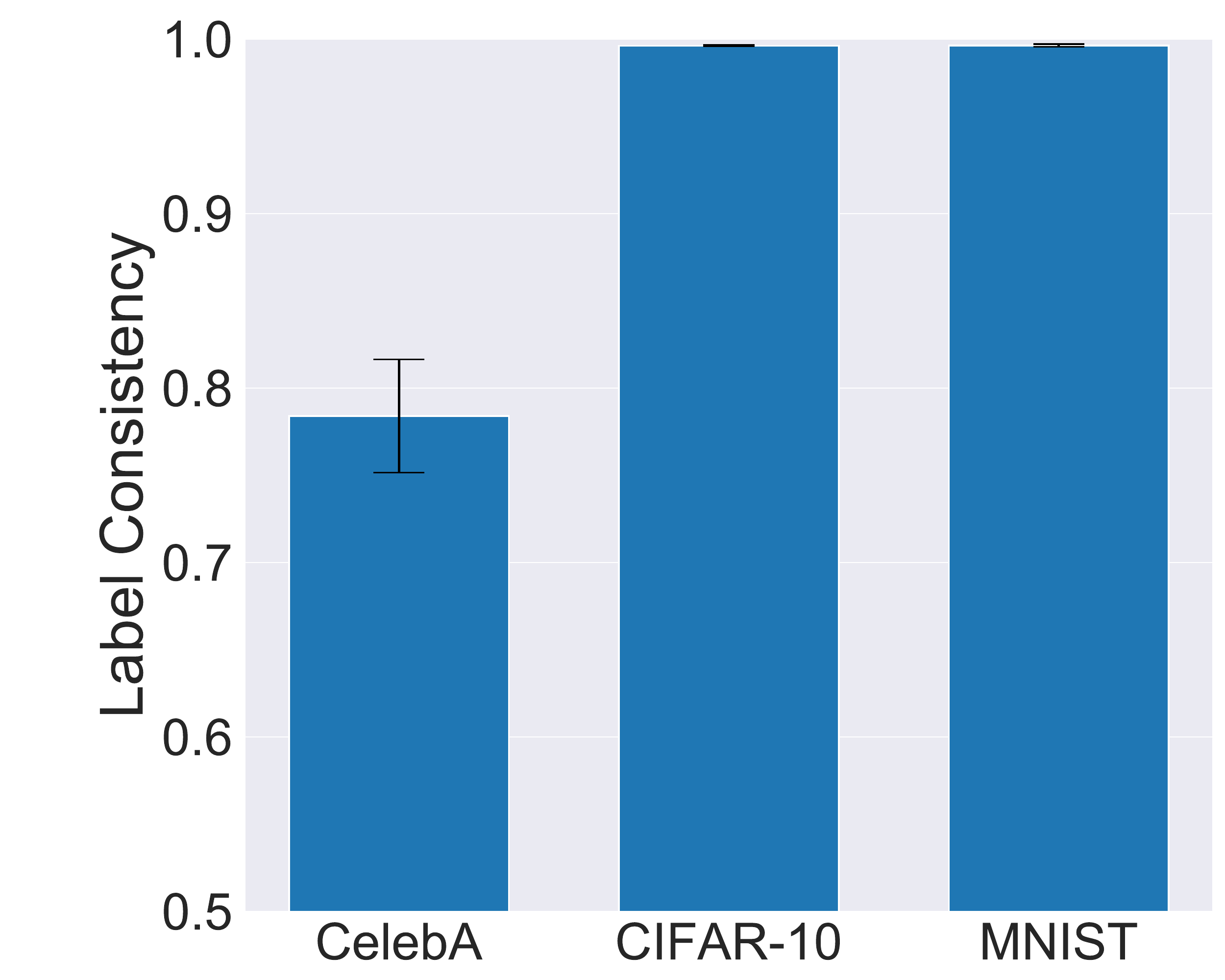}
\caption{Label Consistency}
\label{figure:LC}
\end{subfigure}
\begin{subfigure}{0.4\columnwidth}
\includegraphics[width=1\columnwidth]{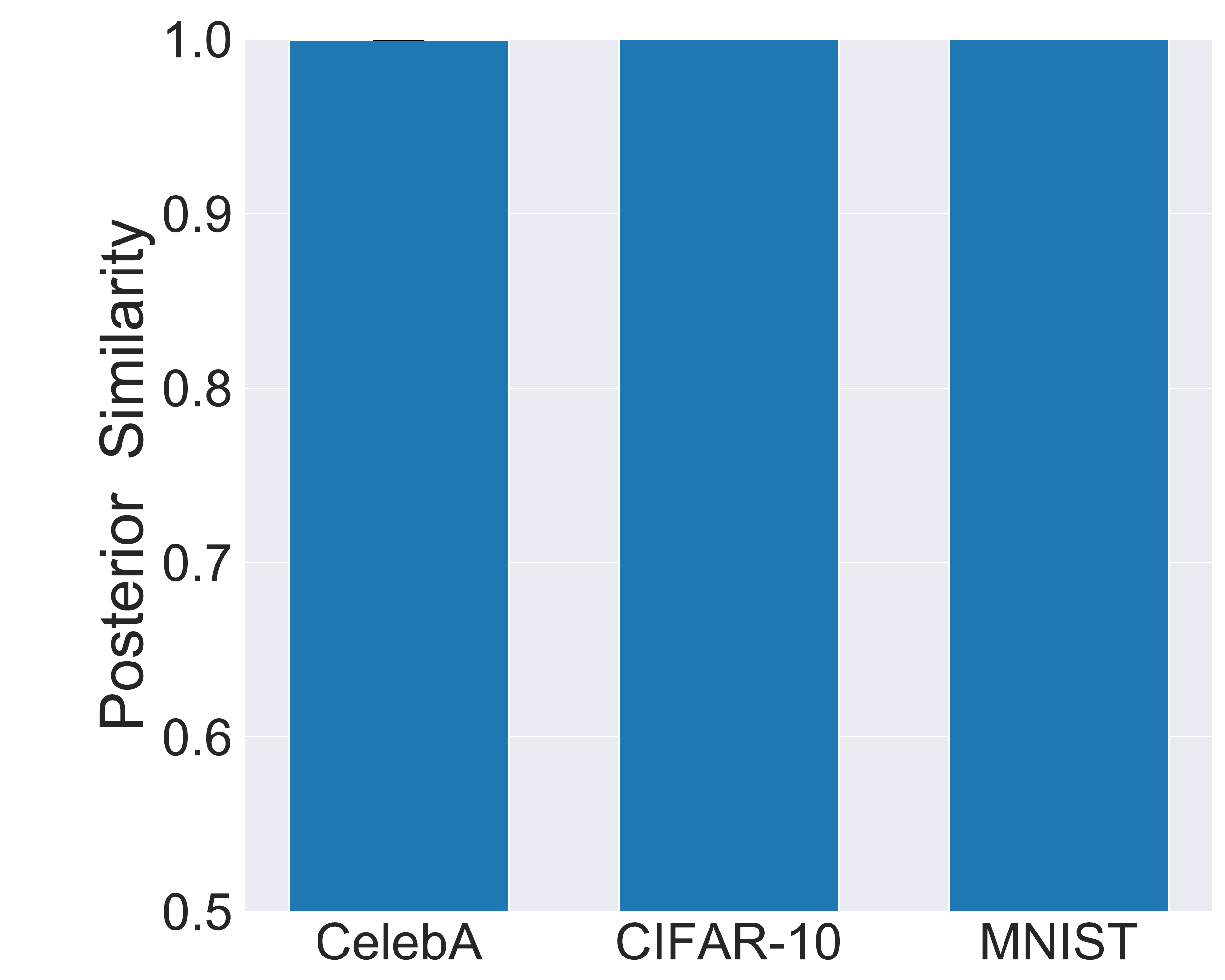}
\caption{Posterior Similarity}
\label{figure:cosineSim}
\end{subfigure}
\begin{subfigure}{0.4\columnwidth}
\includegraphics[width=1\columnwidth]{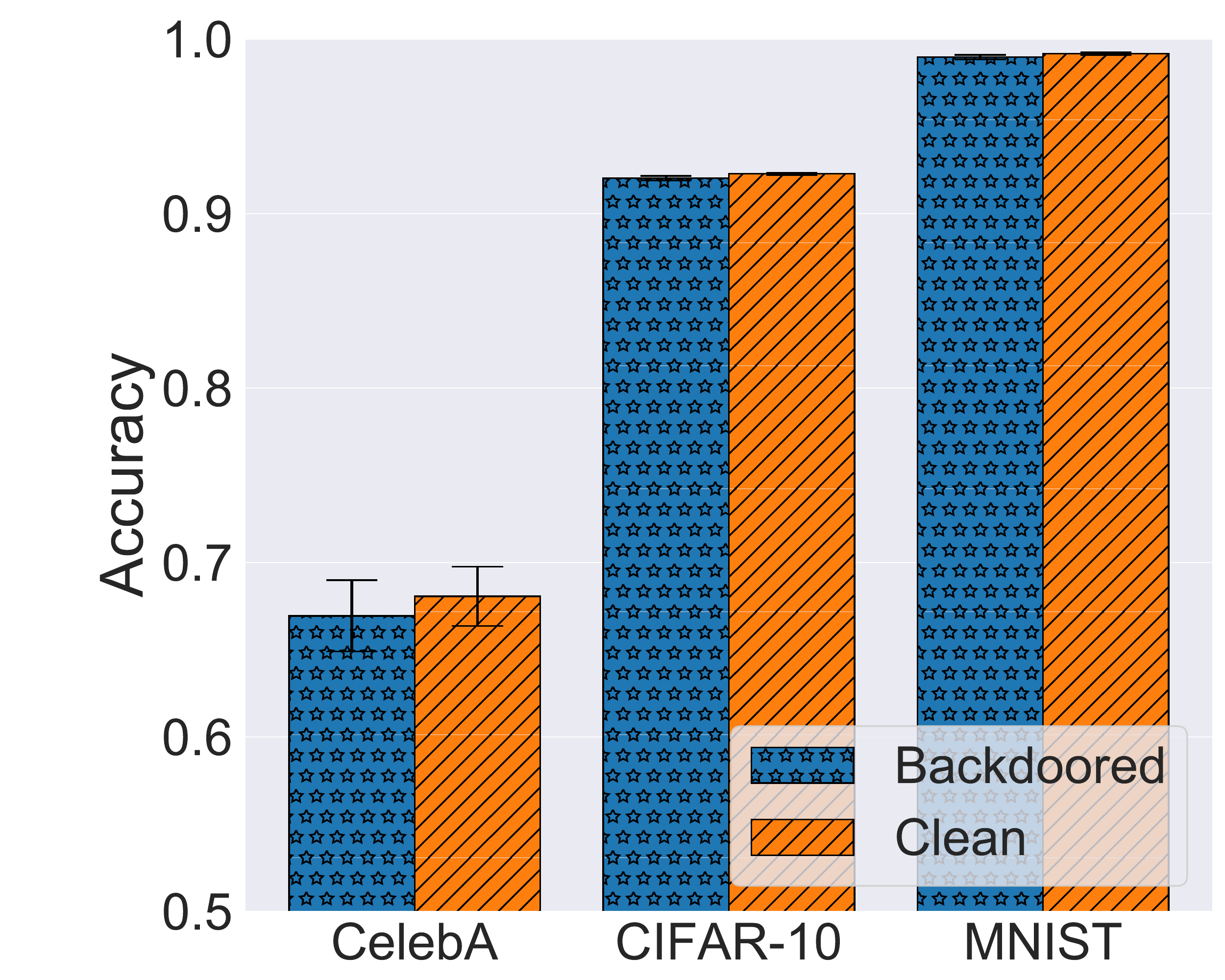}
\caption{Model Utility}
\label{figure:modelUtil}
\end{subfigure}
\caption{
Evaluation of the triggerless backdoor attack when setting the number of queries to 5,000, on the MNIST, CIFAR-10 and CelebA datasets. 
The x-axis represents the different datasets and the y-axis represents the attack success rate (\autoref{figure:ASR}), label consistency (\autoref{figure:LC}), posterior similarity (\autoref{figure:cosineSim}), and accuracy on the clean testing dataset (\autoref{figure:modelUtil}).
}
\label{figure:mainAttack}
\end{figure*} 

We now evaluate our triggerless backdoor attack.
We use all three datasets in our experiments and split each of them into training and testing datasets as follows:
For MNIST and CIFAR-10, we use the default training and testing datasets.
For CelebA, we randomly sample 10,000 sample for both training and testing datasets.
Then, we follow~\autoref{section:meth} to implement our triggerless backdoor in the target models.
We set the target neurons to be a single neuron in the second to last layer. 

For all datasets, we set the number of epochs to train the target models to 50 and train 10 different models for each dataset.
After training, we set the dropout rate to 0.1\% and set the number of queries to 5,000.
\autoref{figure:mainAttack} plots the evaluation results (both mean and standard deviation) for all three datasets.

As \autoref{figure:ASR} shows, our attacks are able to achieve almost a perfect success rate (100\%) on all the three datasets.
It is important to recap that we calculate the attack success rate with respect to the number of queries, i.e., we query the input multiple times and consider the attack successful if one of the outputs is the target label.
Similarly, our attacks achieve a perfect posterior similarity (1) for all three datasets (\autoref{figure:cosineSim}).

However, for label consistency (\autoref{figure:LC}), the result on CelebA is only $0.78$, unlike the results on CIFAR-10 and MNIST both of which have a label consistency of 1.
This is because label consistency is a more strict evaluation metric, i.e., for each input, as long as there is one different label, we consider its label consistency to be 0.
Intuitively, our results for the CelebA dataset shows that the model's outputs are similar, however, the target model seldomly tends to predict a different output label.
To validate this, we repeat the label consistency experiment for the CelebA dataset while counting how many times the input is predicted to more than 2 labels, i.e., the target label and the original prediction.
As expected, the average number of times the input is predicted to another label is only 23.4 (for 5,000 queries).
In other words, there is less than 0.5\% chance that an input is predicted to a third label.

Finally, for model utility (\autoref{figure:modelUtil}), our models are able to achieve a similar performance as clean models.
For instance, our backdoored models achieve 92\%, 67\%, and 99\% accuracy for CIFAR-10, CelebA, and MNIST, respectively, which is only about 0.2\%, 1.1\%, and 0.2\% lower than the clean models.

These results show the efficacy of our triggerless backdoor attack on all three datasets.
Moreover, it is important to note that one of the most important advantages of our attack is that it does not modify the inputs dissimilar to other state-of-the-art backdoor attacks~\cite{GDG17,SWBMZ20,LMALZWZ19}.

\begin{figure}[!t]
\centering
\includegraphics[width=0.9\columnwidth]{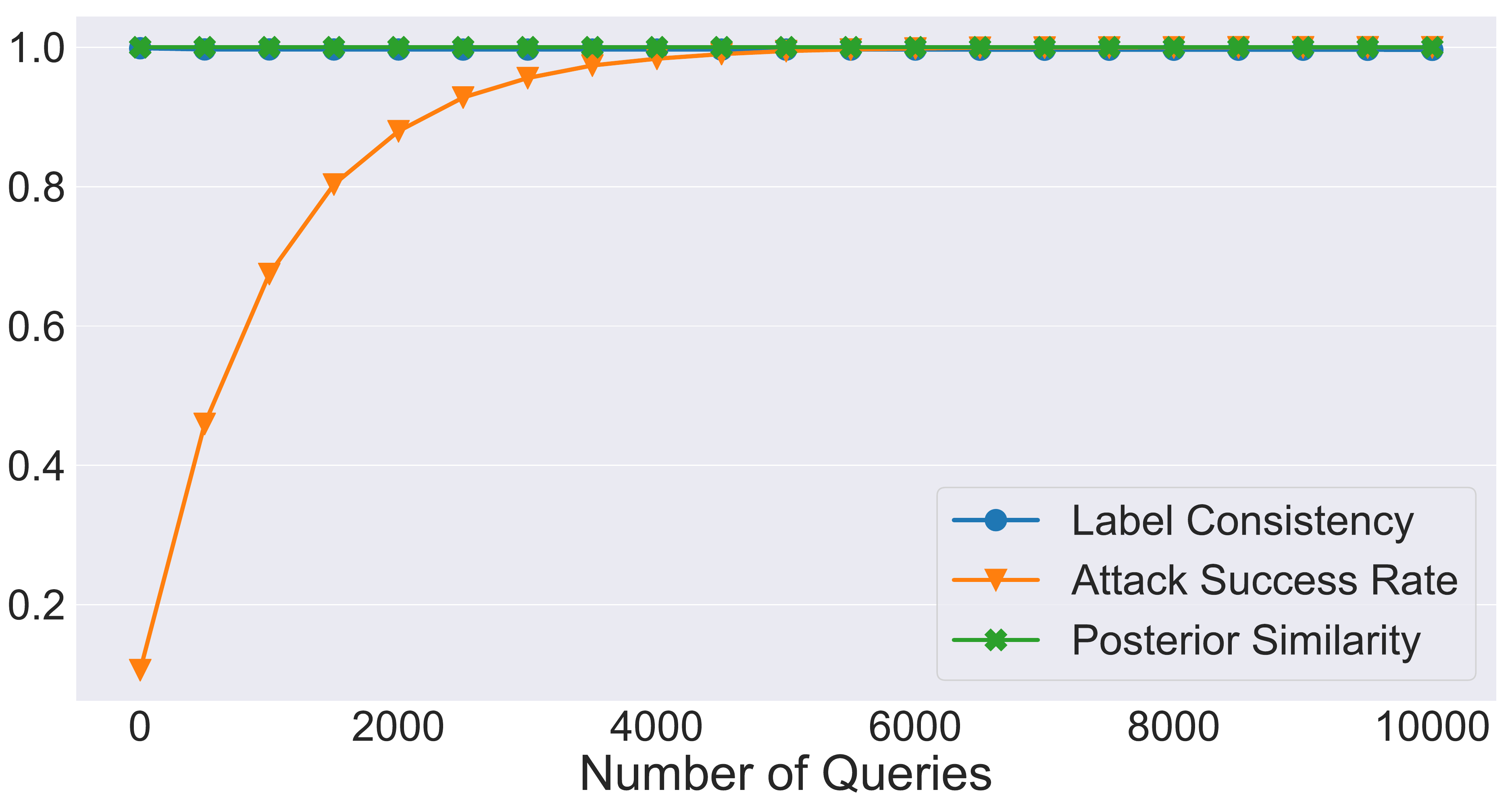}
\caption{
Evaluation of varying the number of queries on the CIFAR-10 dataset. 
The x-axis represents the number of queries and the y-axis represents the different metrics values.
}
\label{figure:numberOfQueries}
\end{figure}

\begin{figure*}[!t]
\centering
\begin{subfigure}{0.9\columnwidth}
\includegraphics[width=1\columnwidth]{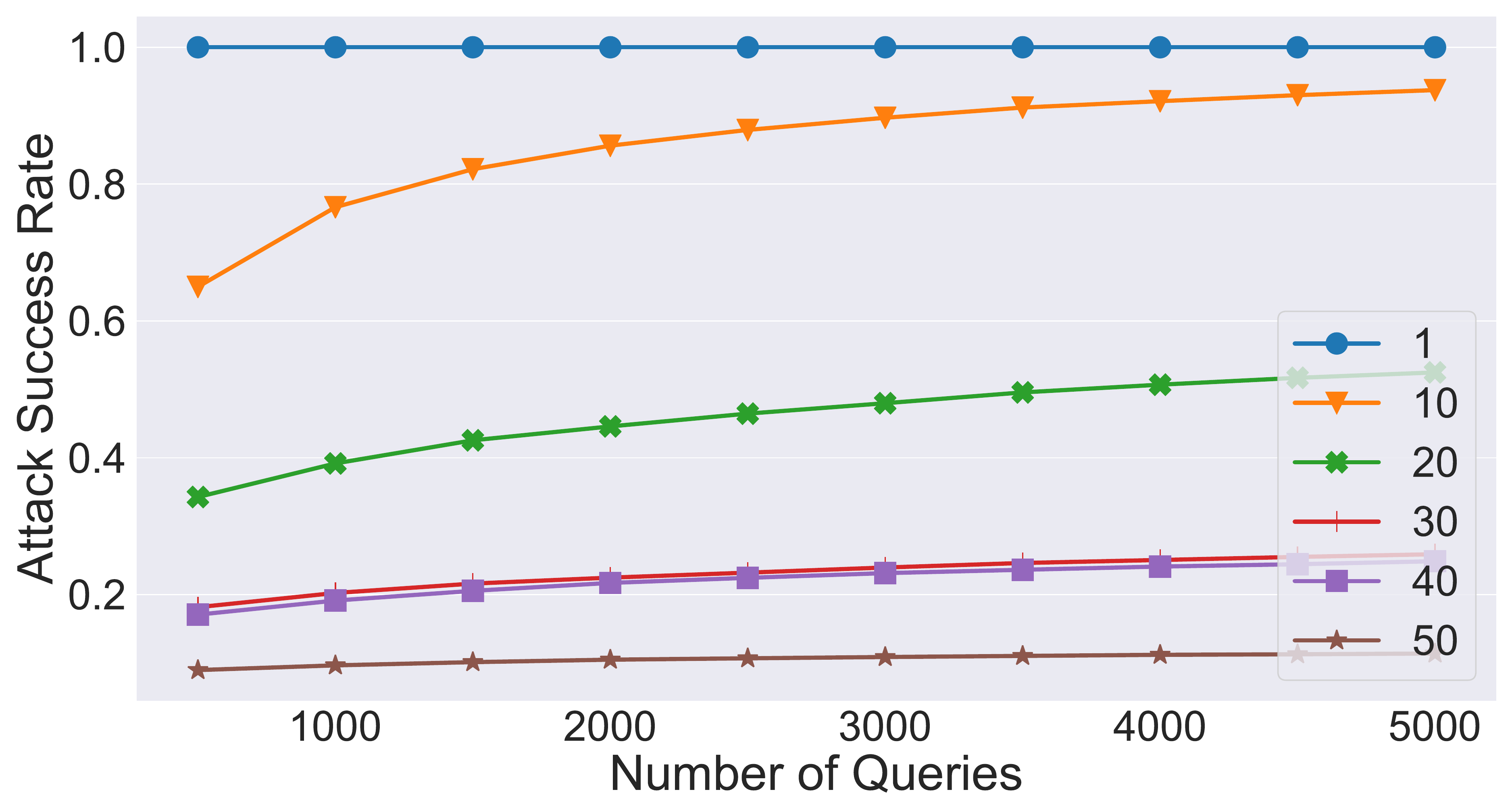}
\caption{Attack Success Rate}
\label{figure:numberOFNeuronsASR}
\end{subfigure}
\quad
\begin{subfigure}{0.9\columnwidth}
\includegraphics[width=1\columnwidth]{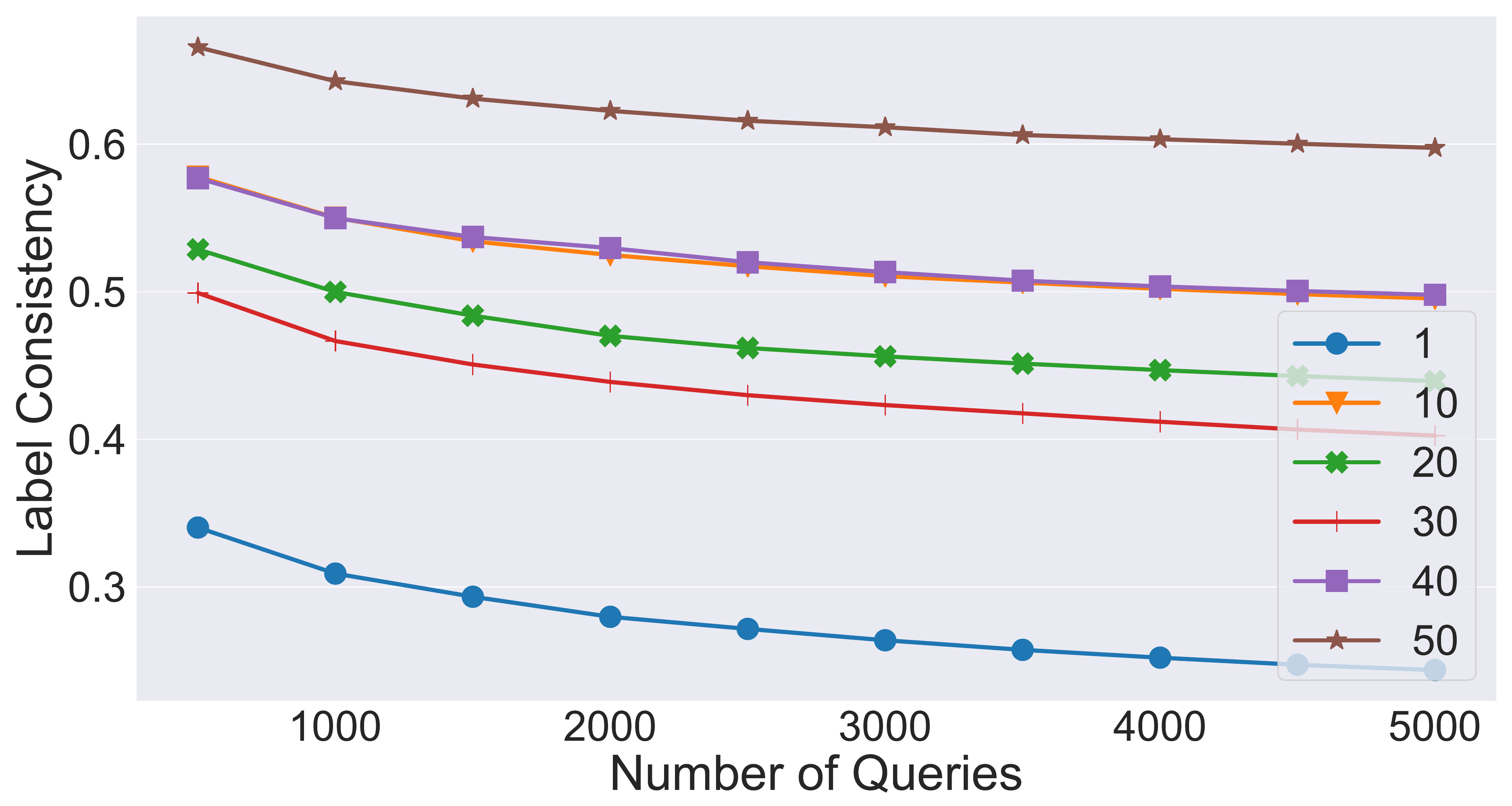}
\caption{Label Consistency}
\label{figure:numberOFNeuronsLC}
\end{subfigure}
\caption{
Evaluation of varying the number of target neurons using the CelebA dataset. 
The x-axis represents the number of quries and the y-axis represents the attack success rate (\autoref{figure:numberOFNeuronsASR}) and label consistency (\autoref{figure:numberOFNeuronsLC}).
}
\label{figure:numberOFNeurons}
\end{figure*} 

% ----------------------------------
\subsection{Hyperparameters Evaluation}
% ----------------------------------

We now evaluate the effect of varying the hyperparameters of our triggerless backdoor attack.
For all of our experiments in this section, we follow the previously introduced evaluation settings (\autoref{section:evalSettings}) with some exceptions that we state for each experiment separately.

\mypara{Number of Queries}
First, we explore the effect of varying the number of queries on our attack.
We use the CIFAR-10 dataset and fix the other experimental settings.
We try from 1 query to 10,000 queries with a step of 500 and plot the results in~\autoref{figure:numberOfQueries}.

As expected, a larger number of queries result in a better attack success rate.
For instance, our triggerless backdoor attack achieves approximately 46\%, 80\%, and 92\% attack success rate for 500, 1,500, and 2,500 queries, respectively.
For both, the label consistency and posterior similarity the performance stays consistent even with a larger number of queries.
For instance, the difference between the label consistency for 500 and 10,000 queries is less than 0.06\%, which demonstrates the robustness of our attack.

\mypara{Number of Target Neurons}
Second, we explore the effect of increasing the number of target neurons, i.e., the neurons that need to be dropped for the backdoor to be activated.
We use the CelebA dataset for this experiment.
We consider models with different range of target neurons, including 1,10, 20, and 50.

With an increase in the number of target neurons, we need to increase the dropout rate as well since the previously used dropout rate (0.1\%) does not drop enough neurons.
Therefore, we set the dropout rate to 10\% for our experiments. 

We evaluate the backdoored models with a different number of queries and plot the results in~\autoref{figure:numberOFNeurons}.
First, we compare the attack success rate of the models with a different number of target neurons.
As expected, fewer target neurons lead to a higher possibility of triggering the backdoor.
For instance, backdooring a model with 1 target neuron can achieve perfect a success rate with less than 500 queries, while a model with 50 target neurons can merely get a 15\% attack success rate with 5,000 queries.

Second, \autoref{figure:numberOFNeuronsLC} compares the label consistency of the models.
Contrary to the attack success rate, label consistency increases with the larger number of target neurons.
The maximum label consistency score that a model with a single target neuron achieves is about 35\% -- note that here we are using a dropout rate of 10\% but \autoref{figure:LC} uses 0.1\%, hence the difference in performance -- which is less than the half of what a model with 50 target neurons achieve.
The gap between the scores of both models even increases with a larger number of queries.
We observe similar behavior for the posterior similarity but with smaller performance gap between different models.

Finally, for the model utility of different models.
As expected, a larger number of target neurons make the model more stable as to trigger the backdoor more neurons are needed.
For instance, there is a gap of about 10\% between the performance of the single target neuron and 50 target neurons.
It is important to note that these results are with a dropout rate of 10\%, however, as previously shown, a single target neuron model can achieve better results in term of label consistency, posterior similarity, and model utility with a lower dropout rate but at the expense of more number of queries to achieve a perfect attack success rate. 

\begin{figure}[!t]
\centering
\includegraphics[width=0.9\columnwidth]{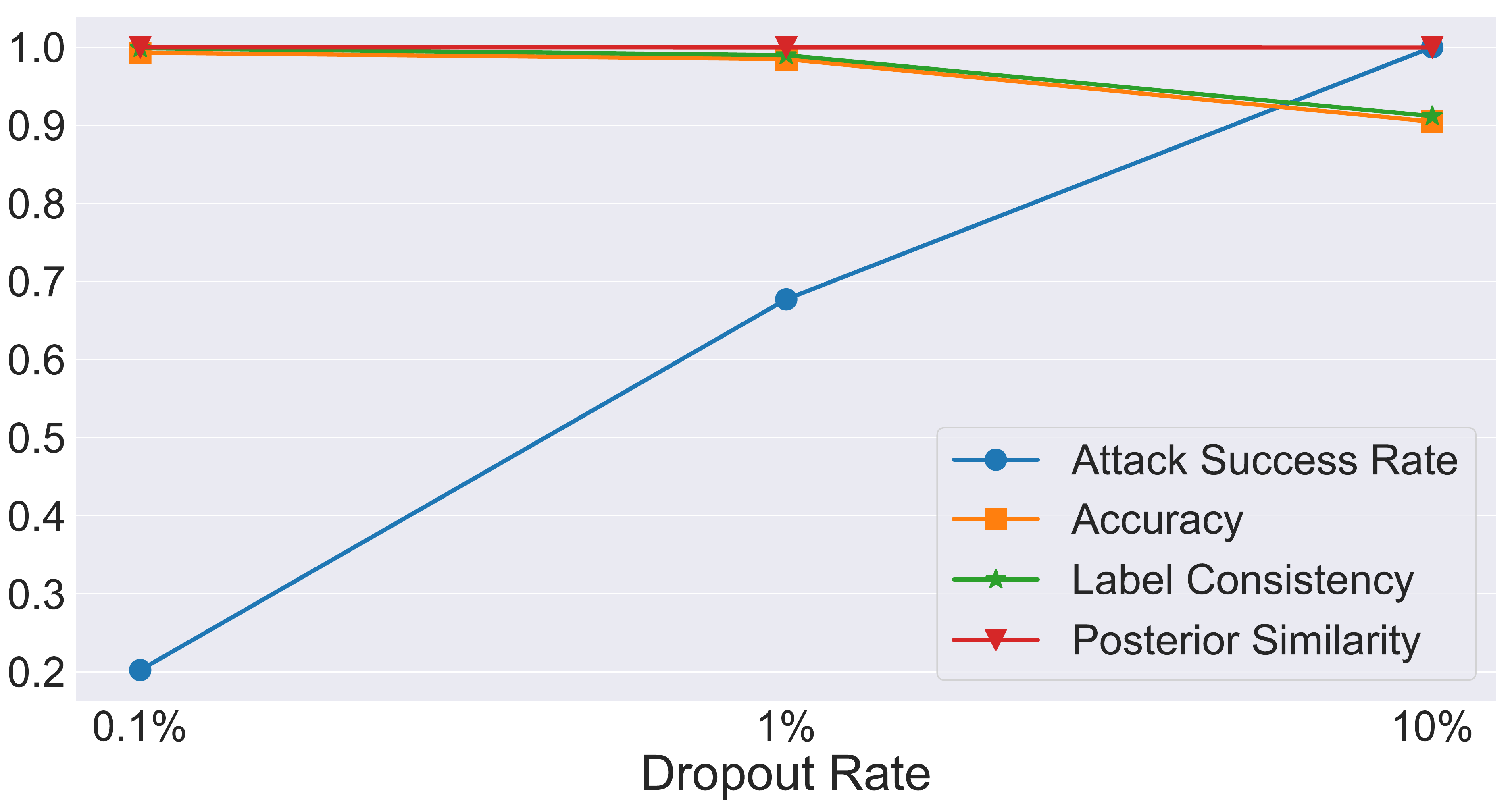}
\caption{
Evaluation of varying the dropout rate while prediction using the MNIST dataset. 
The x-axis represents the dropout rate, and the y-axis represents the different metrics scores.
}
\label{figure:multiDropout}
\end{figure}

\mypara{Dropout Rate}
Third, we explore the effect of using different dropout rates for prediction.
We use the MNIST dataset for this experiment.
We try different dropout rates including 0.1\%, 1\%, and 10\%, and set the number of queries to 100.
\autoref{figure:multiDropout} depicts the result.

Both model utility and label consistency decrease with larger dropout rates.
Posterior similarity also drops, however, with negligible quantity, i.e., it drops by less than 0.01\%.
Moreover, the attack success rate increases significantly with a higher dropout rate.
For instance, using 100 queries can already achieve a 100\% attack success rate when the dropout rate is 10\% compared to only 20\% when the dropout rate is 0.01\%.

\begin{figure}[!t]
\centering
\includegraphics[width=0.9\columnwidth]{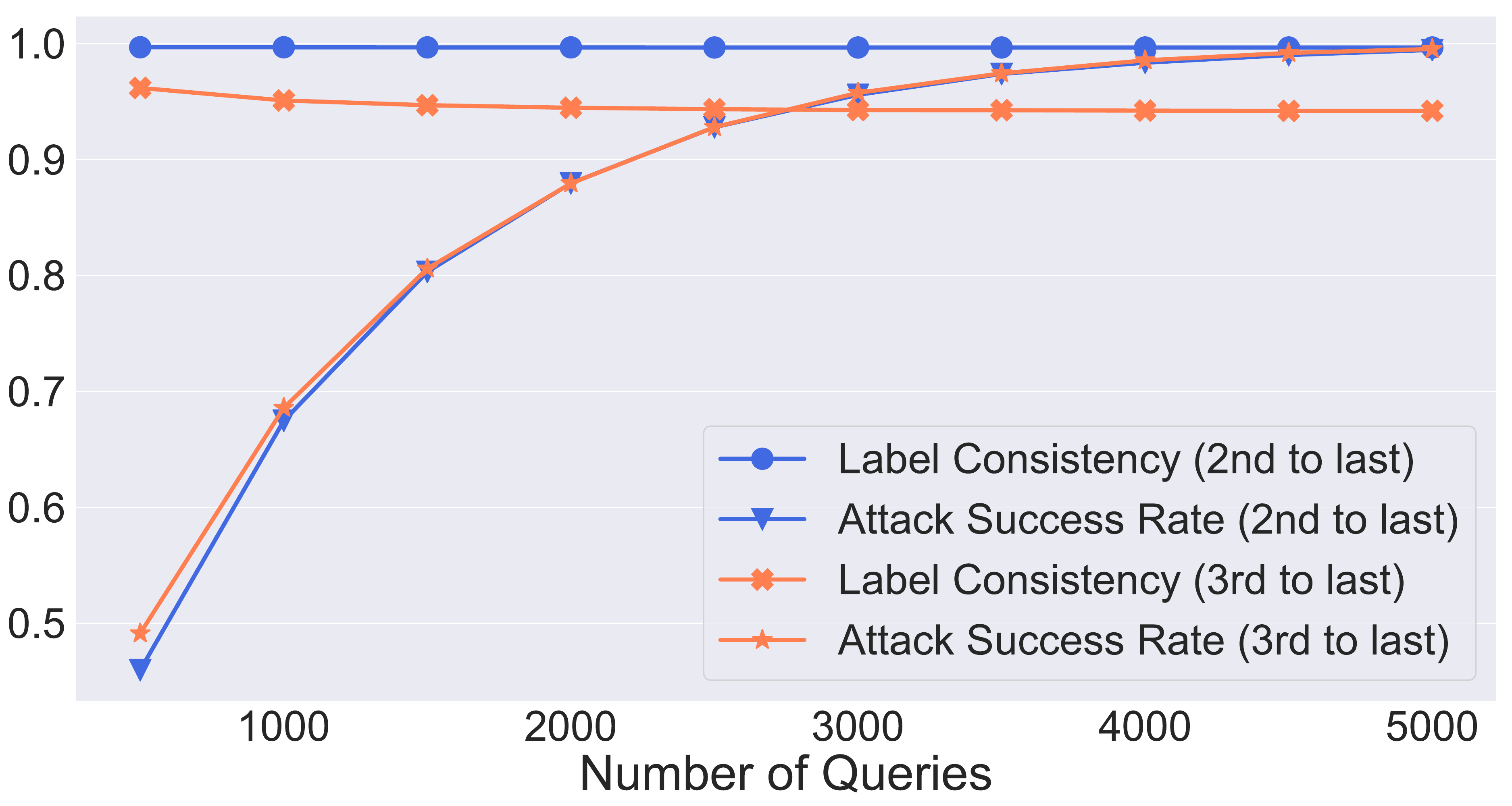}
\caption{
Evaluation of using different layers for the target neuron using the CIFAR-10 dataset.
The x-axis represents the number of queries, and the y-axis represents the different metrics.
}
\label{figure:diffLayer}
\end{figure}

\mypara{Different Target Layer}
For all the previous experiments, we consider the second to last layer as the target layer.
We now investigate whether using different layers for the target neurons can influence our attack.
We use the CIFAR-10 dataset to train a triggerless backdoored model with a single target neuron in the first fully connected layer, i.e., the third to last layer.
We compare the performance of the trained model with the one previously used in~\autoref{section:evalMainAttack}, i.e., the target neuron is in the second to last layer.
We plot the comparison of both models in~\autoref{figure:diffLayer} for a different number of queries using the CelebA dataset.

As the figure shows, both models have a small performance gap when considering the attack success rate, e.g., both are able to achieve 100\% attack success rate at about 5,000 queries.
However, for label consistency, there is a larger gap between the two models.
Using the second to last layer for the target neuron achieves a better performance than the other one. 
This is expected as the last layers have a more direct effect on the final predicted label, i.e., it is the input to the last layer which performs final step of prediction.

% ----------------------------------
\section{Conclusion}
% ----------------------------------

Backdoor attacks against deep neural networks received a lot of attention recently.
However, all current works implement backdoor attacks by using triggers in the input domain, e.g., using a white or colored square as a trigger, which hinders these attacks from being deployed in the physical world.

In this work, we introduce the first triggerless backdoor attack, where no triggers need to be added to the model inputs.
This type of backdoor has two main advantages. 
First, it can be easily applied in the physical world since inputs are not modified.
Second, it can bypass state-of-the-art defenses mechanisms in this field, which detect backdoors by finding triggers.

Our attack is implemented by associating a set of neurons being dropped out during training with a target label.
The attack will be launched when target labels are dropped again during the prediction phase.
Our evaluation shows that our triggerless backdoor attack indeed performs as expected and can easily achieve a perfect attack success rate with a negligible damage to models' utility.
Moreover, we evaluate different hyperparameters of our attack and shows its flexibility being adapted to various use cases.
For instance, the adversary can easily control how often the model triggers the backdoor behavior by adapting the dropout rate.

% ----------------------------------
\balance
\bibliography{normal_generated_py3}
\bibliographystyle{plain}
% ----------------------------------

\end{document}